\documentclass[lettersize,journal]{IEEEtran}

\usepackage{amsmath,amsfonts,amsthm}
\usepackage{array}
\usepackage[font=footnotesize, labelfont=bf, labelsep=period]{caption}
\usepackage[caption=false,font=scriptsize,labelfont=bf]{subfig}
\usepackage[font=scriptsize]{subcaption}

\usepackage{url}
\usepackage{verbatim}
\usepackage{graphicx}
\usepackage{cite}
\usepackage{xcolor}
\usepackage{siunitx}
\usepackage{algorithm}
\usepackage{algpseudocode}
\usepackage{amsmath}
\usepackage{booktabs}
\usepackage{multirow}
\usepackage[acronym]{glossaries}
\usepackage{makecell}

\newcommand{\tgamma}{\widetilde{\gamma}}
\newcommand{\gammaogd}{\tgamma^{\text{\scriptsize OGD}}}
\newcommand{\gammahb}{\tgamma^{\text{\scriptsize HB}}}
\newcommand{\gammanag}{\tgamma^{\text{\scriptsize NAG}}}

\newcommand{\cqi}{\text{\small \gls{CQI}}}
\newcommand{\mcs}{u}

\newcommand{\bler}{\text{\small\textsc{BLER}}}

\newcommand{\multiline}[1]{%
	\parbox[t]{\dimexpr\linewidth-\algorithmicindent}{#1\strut}%
}

\newacronym{SINR}{SINR}{signal-to-interference-plus-noise ratio}
\newacronym{CQI}{CQI}{channel quality indicator}
\newacronym{UE}{UE}{user equipment}
\newacronym{MCS}{MCS}{modulation and coding scheme}
\newacronym{BLER}{BLER}{block error rate}
\newacronym{OCO}{OCO}{online convex optimization}
\newacronym{NAG}{NAG}{Nesterov's accelerated gradient}
\newacronym{HB}{HB}{heavy-ball}
\newacronym{OGD}{OGD}{online gradient descent}
\newacronym{OLLA}{OLLA}{outer-loop link-adaptation}
\newacronym{LTS}{LTS}{latent Thompson sampling}
\newacronym{CBS}{CBS}{code block size}
\newacronym{BCE}{BCE}{binary cross-entropy}
\newacronym{SoA}{SoA}{state-of-the-art}
\newacronym{FS}{FS}{Fixed-Share}
\newacronym{FTL}{FTL}{Follow-the-Leader}
\newacronym{HARQ}{HARQ}{hybrid automatic repeat request}
\newacronym{BS}{BS}{base station}
\newacronym{LA}{LA}{link adaptation}
\newacronym{DL}{DL}{downlink}

\begin{document}

\title{\huge SINR Estimation under Limited Feedback\\via Online Convex Optimization}

\author{
		Lorenzo Maggi,
        Boris Bonev,
        Reinhard Wiesmayr,
		Sebastian Cammerer,
		Alexander Keller
		\thanks{
			Lorenzo Maggi, Boris Bonev, Sebastian Cammerer, and Alexander Keller are with NVIDIA.
            Reinhard Wiesmayr is with ETH Zürich (Switzerland). His work was conducted during an internship at NVIDIA.
            Emails: \{lmaggi, bbonev,scammerer,akeller\}@nvidia.com, wiesmayr@iis.ee.ethz.ch.\\
			This work has been submitted to the IEEE for possible publication. Copyright may be transferred without notice, after which this version may no longer be accessible.}
}


\maketitle

\begin{abstract}
We introduce a novel \gls{OCO} framework to estimate the user's \gls{SINR} from ACK/NACK feedback, \gls{CQI} reports, and previously selected \gls{MCS} values. 
Specifically, the proposed approach minimizes a regularized binary cross-entropy loss using mirror descent enhanced with Nesterov momentum for accelerated \gls{SINR} tracking. Its parameters are tuned online via an expert-advice algorithm, endowing the estimator with continual learning capabilities.
Numerical experiments in ray-traced scenarios show that the proposed method outperforms state-of-the-art schemes in estimation accuracy and adapts robustly to time-varying \gls{SINR} regimes. 
\end{abstract}

\glsresetall

\section{Introduction}

Estimating the \gls{SINR} experienced by the \glspl{UE} at the \gls{BS} is crucial for radio resource management. 
Reliable SINR estimates enable the scheduler to balance spectral efficiency and reliability. 
Conversely, poor estimates can lead to overly aggressive MCS choices, causing retransmissions and increased latency, or overly conservative ones, resulting in reduced throughput.
Unfortunately, accurate \gls{DL} \gls{SINR} estimation at the \gls{BS} is impaired by the nature of uplink feedback. ACK/NACK acknowledgments provide only indications of decoding success or failure, while \gls{CQI} reports may be unreliable due to quantization errors and feedback delay.
Compounding these difficulties is the non-stationary nature of the wireless channel. \Gls{UE} mobility, fading, and interference cause the SINR to evolve in complex patterns. Self-adaptivity is thus a desirable property for an \gls{SINR} estimator to enable both fast adaptation to sudden changes in the \gls{SINR} regime and stable operation during stationary periods.

\textbf{Prior art} on \gls{SINR} estimation from limited feedback is mainly related to adaptive \gls{MCS} schemes. The industry standard is \gls{OLLA} \cite{pedersen2007frequency}. Several variants of \gls{OLLA} have been studied, including stepsize adaptation mechanisms \cite{zhu2023nolla}, reinforcement learning-assisted approaches \cite{kela2022reinforcement}, and customized versions, e.g., for extended reality \cite{paymard2022enhanced}. 
The work \cite{saxena2021reinforcement} proposes \gls{LTS} to infer the \gls{SINR}, while \cite{gautam2025cqi} uses a Kalman filter to infer the interference pattern from \gls{CQI}. 

\textbf{Our contributions} are three-fold:
\begin{itemize}
    \item [i)] We introduce a novel framework for SINR estimation from ACK/NACK and CQI reports, grounded in \gls{OCO}.
    
    \item [ii)] We derive a mirror-descent \gls{SINR} estimator augmented with Nesterov momentum enabling accelerated \gls{SINR} tracking. 
    Intuitively, the estimate is adjusted according to the ``surprise'' of receiving an ACK/NACK, and fused with the CQI estimate via a convex combination.
    
    \item[iii)] We show how the estimator's parameters can be automatically tuned online via an expert-advice algorithm.
\end{itemize} 
This work extends \cite{wiesmayr2025salad}, relying on a simpler \gls{SINR} estimator which is not based on \gls{OCO}, does not exploit CQI report and has no self-optimization properties.

\section{Problem statement} \label{sec:pb_statement}

\textbf{Our scenario} is a 5G NR \gls{DL} communication link between a \gls{BS} and a \gls{UE}. 
Before transmission in each slot $t\in\{1,2,\dots\}$, the \gls{BS} first estimates the \gls{SINR} $\gamma_t$ \SI{}{[\dB]} experienced by the \gls{UE} in slot $t$ as $\tgamma_t$. 
Then, the \gls{BS} selects the \gls{MCS} $\mcs_t$, the code block size $b_t$, and schedules radio resources for communication in slot $t$. 
The design of the \gls{MCS} selection and scheduling policy is outside the scope of this work; the selected \glspl{MCS} are treated as observations.
After transmission, the \gls{BS} receives two types of feedback from the \gls{UE}:
\begin{itemize}
	\item ACK ($y_t=0$) or NACK ($y_t=1$), whether the packet was received by the \gls{UE} correctly or not, respectively. 
	\item $\cqi_t$, typically an integer between 1 and 15, being correlated with the true \gls{SINR} and available only intermittently. 
\end{itemize}

\textbf{Our goal} is to estimate the \gls{SINR} by minimizing the mismatch between candidate \gls{SINR} values and the ACK/NACK and \gls{CQI} feedback, quantified at each slot $t$ by two separate terms (defined formally in Sec.~\ref{sec:sinr_est}):
\begin{itemize}
	\item The loss $\ell_t(\tgamma_t):=\ell_t(\tgamma_t,y_t)$, measuring the discrepancy between the observed $y_t$ and the \gls{SINR} estimate $\tgamma_t$;
	\item The regularizer $\Omega_t(\tgamma_t):=\Omega_t(\tgamma_t,\cqi_t)$, measuring the proximity of the estimate $\tgamma_t$ to the \gls{CQI} estimate. If \gls{CQI} is not reported in slot $t$, then $\Omega_t(\tgamma_t):=0$. 
\end{itemize}
Specifically, we estimate the \gls{SINR} by minimizing the sum of the regularized losses across slots, as
\begin{equation} \label{eq:obj}
	\min_{\{\tgamma_t\}_{t\ge 1}} \sum_{t\ge 1} \ell_t(\tgamma_t) + \Omega_t(\tgamma_t).
\end{equation}

We observe that \eqref{eq:obj} can be solved exactly only once the communication is terminated. 
Yet, to schedule resources, the \gls{BS} must estimate the \gls{SINR} online, in each slot $t$ and relying only on past feedback $\{y_i,\cqi_i\}_{i<t}$. In this case, the \gls{SINR} estimation procedure adheres to the online optimization paradigm, outlined in Alg.~\ref{alg:online_opt} and studied in the next section.

\begin{algorithm}
	\caption{Online optimization for \gls{SINR} estimation} \label{alg:online_opt}
	\begin{algorithmic}[1]
		\State \textbf{Assumption}: Gradients $\nabla \ell_t,\nabla\Omega_t$ are known, $\forall\, t\ge 1$
        \State Initialize \gls{SINR} prediction $\tgamma_1$ for slot 1
		\For{$t\ge 2$}
		\State \multiline{Observe selected \gls{MCS} $\mcs_{t-1}$,  ACK/NACK feedback  $y_{t-1}$ and, optionally, reported $\cqi_{t-1}$ for slot $t-1$}
		\State Incur the penalty $\ell_{t-1}(\tgamma_{t-1})+\Omega_{t-1}(\tgamma_{t-1})$
		\State Estimate the \gls{SINR} in slot $t$ as $\tgamma_t$
		\EndFor
	\end{algorithmic}
\end{algorithm}

\section{\gls{SINR} estimation method} \label{sec:sinr_est}

We next present our online \gls{SINR} estimation method. 
For the sake of clarity, we first consider ACK/NACK-only feedback (Sec.~\ref{sec:ocoacknack}) and then incorporate CQI reports (Sec.~\ref{sec:cqi}).

\subsection{\gls{SINR} estimation from ACK/NACK feedback} \label{sec:ocoacknack}

We first consider the case with ACK/NACK feedback only, discarding \gls{CQI} and setting the regularizer $\Omega_t:=0$, $\forall\,t\ge 1$. 

Let us introduce the \gls{BLER} as
\begin{equation}
    \bler_t(\tgamma) := \Pr(y_t=1|\gamma_t=\tgamma,\mcs_t,b_t), \quad \forall\, t\ge 1
\end{equation}
which is approximated as a sigmoid, as in, e.g., \cite{wiesmayr2025salad,carreras2018link},
\begin{align}
	\bler_t(\tgamma) \approx & \, 1 - \sigma_t(\tgamma) \label{eq:approx} \\
	:= & \, 1 - \left(1 + e^{-\frac{\tgamma-c_t}{s_t}}\right)^{-1}, \quad \forall\, \tgamma\in \mathbb{R}, \, t\ge 1. 
\end{align}
where the center $c_t\!:=\!c(\mcs_t,b_t)$ and scale\footnote{As shown in Sec.~\ref{sec:results}, our method is robust to uncertainties in the sigmoid center $c_t$ and scale $s_t$.} $s_t\!:=\!s(\mcs_t,b_t)$ depend on the \gls{MCS} $\mcs_t$ and \gls{CBS} $b_t$. 

We define the \textbf{loss} $\ell_t$ in slot $t\ge 1$ as the \textbf{\gls{BCE}} between the ACK/NACK feedback $y_t$ and its likelihood, i.e., the estimated \gls{BLER}, 
\begin{equation}
\ell_t(\tgamma_t) \! := \! - y_t \!\log \!\bler_t(\tgamma_t) - (1 - y_t) \!\log \!\left( 1 \!-\! \bler_t(\tgamma_t) \right)\!.
\end{equation}
Under approximation \eqref{eq:approx}, the loss function $\ell_t(\cdot)$ is convex, as desired. 
To estimate the \gls{SINR} in the \gls{OCO} setting, the natural option is \textbf{\gls{OGD}}, which updates the \gls{SINR} estimate along the gradient of the last loss function as
\begin{align}
	\gammaogd_{t} = & \, \gammaogd_{t-1} - \eta \tfrac{\partial}{\partial \tgamma} \ell_{t-1}(\gammaogd_{t-1}), \qquad t\ge 2 \\
	= & \, \gammaogd_{t-1} + \eta s_{t-1}^{-1} \left( \bler_{t-1}(\gammaogd_{t-1}) - y_{t-1} \right) \label{eq:salad} 
\end{align}
where $\eta>0$ is the stepsize and $\gammaogd_1$ is chosen arbitrarily. 

\textbf{Interpretation}. In expression~\eqref{eq:salad}, the term $(\bler_{t-1} - y_{t-1})$ measures the ``surprise'' of observing $y_{t-1}$: if, e.g., an ACK ($y_{t-1}=0$) is observed and $\bler_{t-1}\gg 0$, then $\gammaogd_{t}$ increases significantly, since the \gls{SINR} was underestimated. 

\begin{figure}
	\centering
	\includegraphics[width=\linewidth]{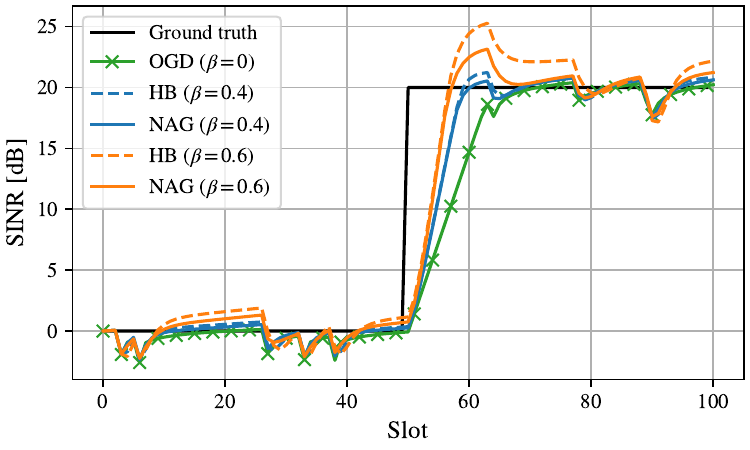}
	\caption{\textbf{\gls{SINR} estimation with momentum}. A comparison among \gls{OGD}, \gls{NAG} and \gls{HB} momentum methods for \gls{SINR} estimation from ACK/NACK. \Gls{NAG} offers a good compromise between tracking speed and stability.} 
	\label{fig:nag_vs_hb}
\end{figure}

\textbf{Adding some inertia} to the gradient descent update can accelerate the \gls{SINR} tracking.
E.g., one can apply the \gls{HB} method \cite{polyak1964some} by adding to the gradient update \eqref{eq:salad} a scaled version of the previous estimate difference $\tgamma_{t-1} - \tgamma_{t-2}$,
\begin{equation}
    \gammahb_{t} = \gammahb_{t-1} + \eta s_{t-1}^{-1} \left( \bler_{t-1}(\gammahb_{t-1}) - y_{t-1} \right) + \beta (\gammahb_{t-1} - \gammahb_{t-2}),
\end{equation}
with $\beta>0$. To understand its behavior, suppose that the \gls{SINR} surges. Upon receiving ACKs, the \gls{BCE} loss gradient has large magnitude---especially if the \gls{MCS} is selected aggressively---due to the resulting ``surprise'' effect. 
The \gls{HB} momentum amplifies this effect, causing a rapid increase in the \gls{SINR} estimate. 
However, \gls{HB} overshoots once the estimate reaches the true value and the first NACKs are received, see Fig.~\ref{fig:nag_vs_hb}.

This instability can be mitigated by the \textbf{\gls{NAG}} method \cite{nesterov1983method}, computing the gradient at the look-ahead position $m_{t-1} = \gammanag_{t-1} + \beta(\gammanag_{t-1} - \gammanag_{t-2})$, i.e., the point where the momentum projects the estimate:
\begin{equation}
    \gammanag_t \!= m_{t-1} \!+ \eta s_{t-1}^{-1} \left( \bler_{t-1}(m_{t-1}) \! - \! y_{t-1} \right)\!, \, \forall t \! \ge \!2.
\end{equation}
Since $m_{t-1}$ is closer to the true value than the current estimate $\gammanag_{t-1}$, an ACK is less surprising than in \gls{HB} when the \gls{SINR} increases. This reduces oscillations as the estimate approaches the target, as illustrated in Fig.~\ref{fig:nag_vs_hb}.

\subsection{Integrating the \gls{CQI} feedback}  \label{sec:cqi}

We now reintroduce the \gls{CQI} feedback and tackle our original problem in Eq.~\eqref{eq:obj}. 
The \gls{CQI} is mapped to an \gls{SINR} estimate via the function $f(\cdot)$ which can be defined as in \cite[Tables 5.2.2.2-5]{3gpp38214}, else learned from data, e.g., as in \cite{jiang2025ai}. 

We conveniently design the convex regularizer term $\Omega_t$ as the squared distance between the estimate $\tgamma_t$ and $f(\cqi_t)$:
\begin{equation}
	\Omega_t(\tgamma_t) := \tfrac{\lambda}{2} \left( \tgamma_t - f(\cqi_t) \right)^2, \quad \forall\, t\ge 1
\end{equation}
where $\lambda>0$ if the \gls{CQI} is reported for slot $t$, otherwise $\lambda=0$.

The \textbf{mirror descent} method in \cite{duchi2010composite} solves the regularized problem \eqref{eq:obj} online. In each slot $t\ge 2$, it minimizes a linearized version of the loss, $\tfrac{\partial}{\partial \tgamma}\ell_t(\tgamma_{t-1}) \tgamma$, penalized by the squared Euclidean distance\footnote{The formulation in \cite{duchi2010composite} is more general than ours since it considers the Bregman divergence $D_\phi(\tgamma,\tgamma_{t-1})=\phi(\tgamma) - \phi(\tgamma_{t-1}) - \tfrac{\partial}{\partial \tgamma}\phi(\gamma_{t-1})(\tgamma-\tgamma_{t-1})$, which reduces to the squared Euclidean distance if $\phi(\cdot)=\tfrac{1}{2}(\cdot)^2$.} from the current estimate, $\frac{1}{2\eta}(\tgamma - \tgamma_{t-1})^2$, where $\eta>0$, plus the regularizer $\Omega_{t-1}$, as
\begin{equation} \label{eq:mirror_descent}
	\tgamma_{t} = \underset{\tgamma\in\mathbb{R}}{\mathrm{argmin}} \tfrac{\partial}{\partial \tgamma} \ell_{t-1}(\tgamma_{t-1}) \tgamma + \tfrac{1}{2\eta} (\tgamma - \tgamma_{t-1})^2 + \Omega_{t-1}(\tgamma).
\end{equation}
By setting the derivative of \eqref{eq:mirror_descent} relative to $\tgamma$ to zero, we obtain
\begin{equation}  \label{eq:convex_combination} 
	\tgamma_{t} = \alpha f(\cqi_{t-1}) + (1-\alpha) \big(\tgamma_{t-1} - \eta \tfrac{\partial}{\partial \tgamma} \ell_{t-1}(\tgamma_{t-1})\big)
\end{equation}
for $t\ge 2$, where $\alpha := \eta \lambda/(1 + \eta \lambda)\in[0,1]$.

\textbf{Interpretation}. Expression \eqref{eq:convex_combination} has a natural interpretation: the \gls{SINR} estimate $\tgamma_t$ is the convex combination between the estimate from \gls{CQI} report and the \gls{OGD} update---cfr. Eq.~\eqref{eq:salad}---from ACK/NACK feedback. Moreover, the convex coefficient $\alpha$ represents the level of \textbf{reliance on the \gls{CQI} feedback}, which should depend on the quality of the mapping $f$ (see Sec.~\ref{sec:expert_advice}). 
We also observe that~\eqref{eq:convex_combination} can be interpreted as a Kalman filter, where $\gammanag$, $f(\cqi)$, and $\alpha$ act as the model prediction, the noisy observation, and  the Kalman gain, respectively.

To \textbf{reintroduce Nesterov’s momentum} in the presence of \gls{CQI} feedback, the \gls{OGD} term in \eqref{eq:convex_combination} can be replaced with its \gls{NAG} counterpart.
However, this can lead to uncontrolled oscillations, as the \gls{CQI} report can amplify the momentum term excessively.
For this reason, we propose applying momentum only when is \gls{CQI} not reported, and setting $\beta=0$ otherwise. 

Alg.~\ref{alg:sinr_est_cqi} outlines our \gls{SINR} estimation method with fixed parameters $\alpha$, $\beta$, and $\eta$. In the next section, we show how these parameters can be tuned in an online fashion. 

\begin{algorithm}
	\caption{\gls{SINR} estimate from ACK/NACK and \gls{CQI}} \label{alg:sinr_est_cqi}
	\begin{algorithmic}[1]
		\State \textbf{Parameters}:  $\alpha\in[0,1],\beta\ge 0,\eta\ge 0$
        \State Initialize \gls{SINR} predictions $\tgamma_0,\tgamma_1$
		\For{$t\ge 2$}
		
        \State \multiline{Observe selected \gls{MCS} $\mcs_{t-1}$,  ACK/NACK feedback  $y_{t-1}$ and, optionally, reported $\cqi_{t-1}$ for slot $t-1$}
		
        \State Set $\overline{\alpha}=\alpha$ \textbf{if} $\cqi_{t-1}$ is reported, \textbf{else} $\overline{\alpha}=0$

        \State Set $\overline{\beta}=\beta$ \textbf{if} $\cqi_{t-2}$ is reported, \textbf{else} $\overline{\beta}=0$
        
        \State Apply momentum: $m_{t-1} = \tgamma_{t-1} + \overline{\beta}(\tgamma_{t-1} - \tgamma_{t-2})$
		
        \State Estimate the \gls{SINR} from ACK/NACK as\footnotemark
        $$
        \tgamma'_t = m_{t-1} + \eta s_{t-1}^{-1} \left( \bler_{t-1}(m_{t-1}) - y_{t-1} \right)
        $$
        
        \State Integrate the \gls{CQI} and estimate the \gls{SINR} in slot $t$ as 
		$$
			\tgamma_{t} = \overline{\alpha} f(\cqi_{t-1}) + (1-\overline{\alpha}) \tgamma'_t
		$$
		
		\EndFor
	\end{algorithmic}
\end{algorithm}

\footnotetext{To avoid numerical instabilities, we recommend clipping $s_t$ and $\bler_{t-1}$ to, e.g., $[0.5, 2]$ and $[0.01,0.99]$, respectively.}

\begin{figure}[h]
\centering
	\subfloat[\scriptsize \textbf{Scenario 1}: \gls{SINR} evolution]{\includegraphics[width=.48\columnwidth]{"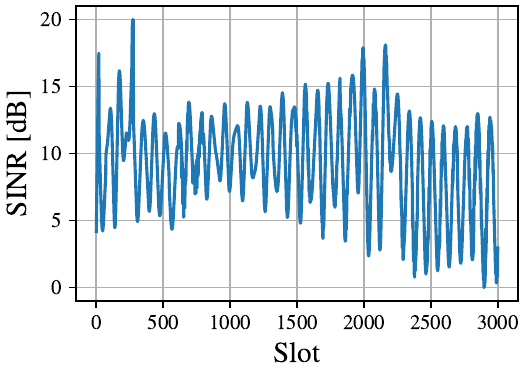"}%
		\label{fig:subfig1}}
	\hfil
	\subfloat[\scriptsize \textbf{Scenario 2}: \gls{SINR} evolution]{\includegraphics[width=.48\columnwidth]{"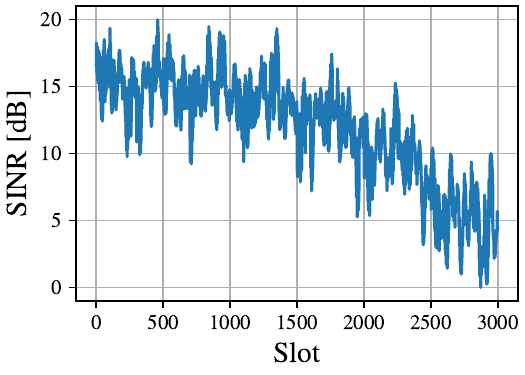"}%
		\label{fig:subfig2}}
    
    \vspace{0pt}

    \subfloat[\scriptsize \textbf{Scenario 1}: RMSE of \gls{SINR} prediction for different $(\eta,\beta)$ and $\alpha=0$.]{\includegraphics[width=.48\columnwidth]{"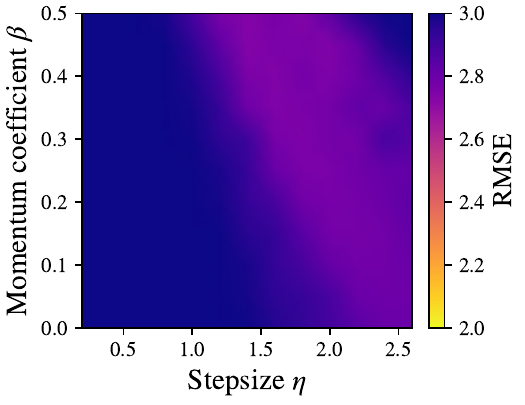"}%
		\label{fig:subfig4}}
	\hfil
	\subfloat[\scriptsize \textbf{Scenario 2}: RMSE of \gls{SINR} prediction for different $(\eta,\beta)$ and $\alpha=0$.]{\includegraphics[width=.48\columnwidth]{"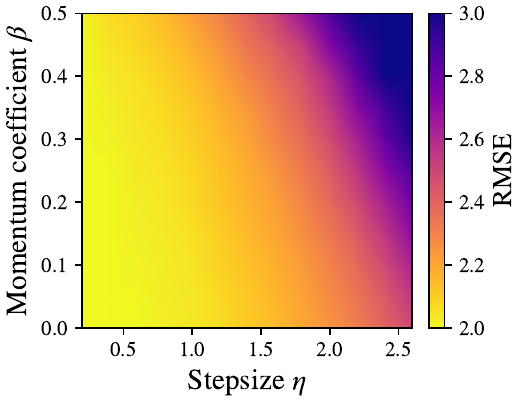"}%
		\label{fig:subfig5}}

    \vspace{0pt}

    \subfloat[\scriptsize \textbf{Scenario 1}: RMSE of \gls{SINR} prediction for different $(\eta,\alpha)$ and $\beta=0$.]{\includegraphics[width=.48\columnwidth]{"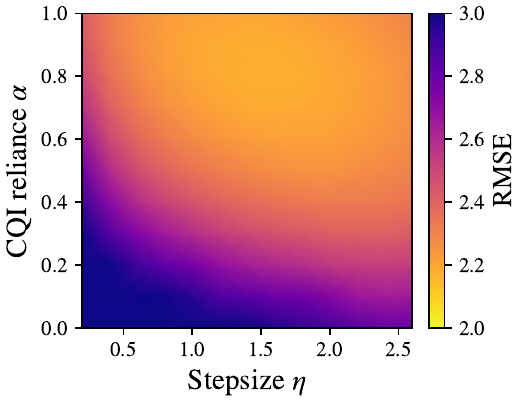"}%
		\label{fig:subfig7}}
	\hfil
	\subfloat[\scriptsize \textbf{Scenario 2}: RMSE of \gls{SINR} prediction for different $(\eta,\alpha)$ and $\beta=0$.]{\includegraphics[width=.48\columnwidth]{"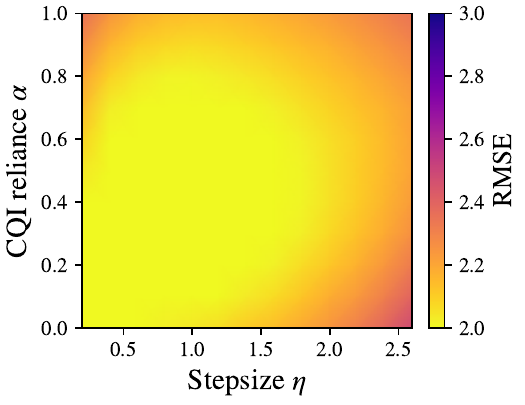"}%
		\label{fig:subfig8}}

    \caption{\textbf{Impact of parameters $\alpha,\beta,\eta$ on \gls{SINR} estimation accuracy}. Root mean square error (RMSE) of the \gls{SINR} prediction evaluated in two ray-traced scenarios across different values of the \gls{CQI} reliance $\alpha$, momentum $\beta$, and stepsize $\eta$. The \gls{MCS} dataset is produced using \gls{OLLA} algorithm \cite{pedersen2007frequency}. The optimal $\alpha,\beta,\eta$ depend on the scenario, motivating the need for online tuning.}
    \label{fig:expert_advice}
    
\end{figure}

\section{Online parameter tuning} \label{sec:expert_advice}

The optimal parameters $\alpha$, $\beta$, and $\eta$ depend on the underlying \gls{SINR} regime. 
A large stepsize $\eta$ enables more effective tracking of rapid \gls{SINR} fluctuations (Fig.~\ref{fig:expert_advice}, scenario~1), while the momentum $\beta$ must balance responsiveness against the risk of overshooting. 
When fluctuations are milder and/or less predictable (Fig.~\ref{fig:expert_advice}, scenario 2), $\eta$ should be reduced to enhance stability. 
Furthermore, the level $\alpha$ of reliance on the \gls{CQI} should reflect both the timeliness of the CQI feedback and the accuracy of the mapping $f(\cdot)$, unknown \emph{a priori}.

The observations above motivate the use of a method that tunes the parameters $\alpha,\beta,\eta$ in an online fashion and adapts to a time-varying SINR regime.
To this end, we adopt the framework of online learning with expert-advice \cite{cesa2006prediction}, consisting of the three following steps: 
\begin{itemize}
    \item[i)] Maintaining multiple parallel instances (``experts'') of Alg.~\ref{alg:sinr_est_cqi}, each with different values of parameters $\alpha,\beta,\eta$.

    \item[ii)] Evaluating the performance of each expert on past data using a loss function $\ell'$---not necessarily the \gls{BCE} loss.

    \item[iii)] Estimating the \gls{SINR} as a weighted average of the expert predictions, whose weights depend on their past losses.
\end{itemize}

\textbf{Continual learning}. When the \gls{SINR} regime changes, the parameters must adapt accordingly.
The \gls{FS} algorithm\footnote{FS was originally proposed in \cite{herbster1998tracking}. Here, we use its simplified version presented in \cite{hazan2016introduction}, Alg. 30.} \cite{herbster1998tracking} can track the time-varying best expert by repeatedly redistributing a fraction of the weights among the experts. This mechanism prevents any irreversible concentration on a single expert and enables continual learning. This is formalized in Alg.~\ref{alg:fixed_share}.

\textbf{Computational complexity}. A single expert (Alg.~\ref{alg:sinr_est_cqi}) estimates the \gls{SINR} in $\mathcal O(1)$ time, while \gls{FS} incurs a linear computational complexity in the number $N$ of experts. In practice, our approach performs well with a small number of experts (around 10), provided their parameters are sufficiently diverse (see Sec.~\ref{sec:results}). Moreover, each expert can also be executed independently and in parallel. 

\begin{algorithm}
	\caption{Self-tuning parameters via Fixed-Share \cite{herbster1998tracking}}\label{alg:fixed_share}
	\begin{algorithmic}[1]
		\State \textbf{Parameters}: $\epsilon>0$, $\mu\in\![0,\!1]$, e.g., $\epsilon\approx 1,\mu\approx 10^{-3}$
		\State Define loss $\ell'$ and $N$ instances of Alg.~\ref{alg:sinr_est_cqi}, i.e., ``experts'', each using different values of $\alpha\in[0,1],\beta\ge 0,\eta\ge 0$
		\State Initialize $w^{(n)}_1=1/N$, $\forall \, n=1,\dots,N$
		
		\For{step $t\ge 1$}
        \State \multiline{Estimate the \gls{SINR} in slot $t$ as $\tgamma_{t}=\sum_{n=1}^N w^{(n)}_t \tgamma_t^{(n)},$ where $\tgamma^{(n)}_t$ is the estimate produced by expert $n$}
        
        \State \multiline{Observe selected \gls{MCS} $\mcs_{t}$,  ACK/NACK feedback  $y_{t}$ and, optionally, reported $\cqi_{t}$ for slot $t$}
		

        \State Update weights as $w^{(n)}_{t+1} = w^{(n)}_t e^{-\epsilon \ell'_t\left(\tgamma^{(n)}_t\right)}, \quad \forall\, n$

        \State Normalize weights as $w^{(n)}_{t+1} \leftarrow \frac{w^{(n)}_{t+1}}{\sum_{n'} w^{(n')}_{t+1}}, \quad \forall\, n$
        \State Redistribute the weights across the experts as 
        \Statex \hspace{30pt} $w^{(n)}_{t+1} \leftarrow (1-\mu) w^{(n)}_{t+1} + \mu \frac{1}{N}, \quad \forall\, n$ 
		
		\EndFor
	\end{algorithmic}
\end{algorithm}

\section{Performance guarantees} \label{sec:theory_guarantees}

\footnote{In this technical section, we assume that the reader has some familiarity with the theory behind \gls{OCO}. We refer to \cite{hazan2016introduction} for an excellent introduction.}The performance of our method in the \textbf{ACK/NACK}-only setting (Sec.~\ref{sec:ocoacknack}) can be evaluated in terms of the \emph{regret} $R_T$, being the difference between the cumulative loss of the algorithm and that of a candidate solution $\{g_t\}_t$ over $T$ slots:
\begin{equation}
	R_T := \sum_{t=1}^T \ell_t(\tgamma_t) - \sum_{t=1}^T \ell_t(g_t).
\end{equation}

Different regret guarantees hold depending on the choice of the candidate solution.
If $g_t$ is the best constant guess, i.e., $g_t:=g$ for all $t$ where $g=\operatorname{arg\,min}_{g'} \sum_{t=1}^T \ell_t(g')$, then we recover the classic static regret.
Since the \gls{SINR} varies over time and our goal is to track it, evaluating the regret with respect to a dynamic candidate is more appropriate.
The regret is typically bounded as a function of the candidate’s variability, quantified by the \emph{path length} $P_T := \sum_{t=2}^T |g_t - g_{t-1}|$.
It is shown in \cite{zinkevich2003online} that \gls{OGD} with a fixed learning rate $\eta$ on the order of $1/\sqrt{T}$ achieves a dynamic regret $\mathcal O(\sqrt{T}(1+P_T))$ against any possible candidate with path length $P_T$.

The acceleration induced by \textbf{momentum} in offline settings does not automatically translate into an improved online regret \cite{alacaoglu2020new}, which typically requires extra structure such as lookahead predictions, as shown in \cite{li2020leveraging}. Yet, momentum can improve \gls{SINR} tracking in practice, as discussed in Sec.~\ref{sec:ocoacknack}.

When \textbf{\gls{CQI}} is considered and composite mirror descent is used (Sec.~\ref{sec:cqi}), the regret also includes the regularizer, i.e., $R_T := \sum_{t=1}^T \ell_t(\tgamma_t) + \Omega_t(\tgamma_t) - \ell_t(g_t) - \Omega_t(g_t)$. 
In this case, one can derive a bound $\mathcal O(\sqrt{T})$ on the static regret \cite{duchi2010composite}.

Finally, we analyze the ability of the \textbf{expert-advice}, \gls{FS}-based Alg.~\ref{alg:fixed_share} to track the best parameter combination $(\alpha,\beta,\eta)$. 
If the identity of the best expert changes $k$ times over $T$ slots, the regret of \gls{FS} on the loss $\ell'$ with respect to an oracle selecting the best expert on each interval is $\mathcal O(k\log(\frac{NT}{k}))$, which is sublinear in $T$ (see \cite{hazan2016introduction}, Ch.~10).
In contrast, an algorithm that converges to the best \emph{fixed} parameters in hindsight incurs linear regret, since it cannot match the best expert on every sub-interval.  This demonstrates the continual learning capability of Alg.~\ref{alg:fixed_share} in non-stationary environments.

\section{Numerical experiments} \label{sec:results}

We evaluate our method using Sionna RT \cite{sionna-rt-tech-report}, an open-source wireless ray-tracer. We consider the multi-cell OpenStreetMap scene from \cite{sionna_rt_radiomaps}, featuring 7 \glspl{BS} equipped with 16 transmit antennas each. The \gls{SINR}, accounting for inter-cell interference from neighboring \glspl{BS}, is computed over 64 subcarriers with \SI{30}{\kHz} spacing. 100 randomly placed \glspl{UE} move along straight-line trajectories at speeds between 3 and 20~\si{\meter\per\second} over 3000 slots. A full-buffer \gls{DL} traffic model is assumed. ACK/NACK and CQI are reported with a delay of 5 slots. 
To evaluate robustness to BLER model mismatch, we intentionally introduce a discrepancy between the algorithm’s internal BLER model and the true link behavior by computing BLER estimates using an incorrect \gls{CBS} ($10^2$ instead of $10^3$).

\begin{figure*}[h!]
    \centering
    \begin{minipage}[t]{0.35\textwidth}
        \vspace{-3.5cm}
        \fontsize{7.8}{9}\selectfont
        \begin{tabular}{@{}lc|cc@{}}
             & \multicolumn{2}{c}{\makecell{SINR estimation RMSE [dB] \\ (20, 50, 80)\% percentile across \glspl{UE}}} \\
            \gls{SoA} algorithm & Baseline & Ours \\
            \hline
            OLLA, $\Delta=2$  & (3.04, 3.53, 4.13) & (\textbf{0.94}, \textbf{1.66}, \textbf{2.45})  \\
            OLLA, $\Delta=1$  & (1.43, 2.04, 2.87) & (\textbf{0.92}, \textbf{1.76}, \textbf{2.48}) \\
            NOLLA, $\Delta=2$  & (1.72,	2.05, 2.87) & (\textbf{0.94}, \textbf{1.65}, \textbf{2.30})  \\
            NOLLA, $\Delta=1$  & (0.94, 1.95, 2.77) & (\textbf{0.83}, \textbf{1.70}, \textbf{2.51})  \\
            LTS & (1.17, 1.73, 2.69) & (\textbf{0.96}, \textbf{1.37}, \textbf{2.54})  \\
            SALAD  & (0.83, 1.62, 2.49) & (\textbf{0.72}, \textbf{1.37}, \textbf{2.11}) \\
        \end{tabular}

        \captionof{table}{\textbf{Comparison with SoA}.
            Datasets of MCS/ACK/NACK feedback are first generated via different \gls{LA} SoA methods on a ray-traced scenario across 100 UEs. Our method is then applied in an online, open-loop fashion to
            estimate SINR, which we compare to the SINR estimates of the SoA baselines.}
        \label{tab:sinr_est_cdf_vs_ollas}
    \end{minipage}\hfill
    \begin{minipage}[t]{0.29\textwidth}
        \centering
        \includegraphics[width=\textwidth]{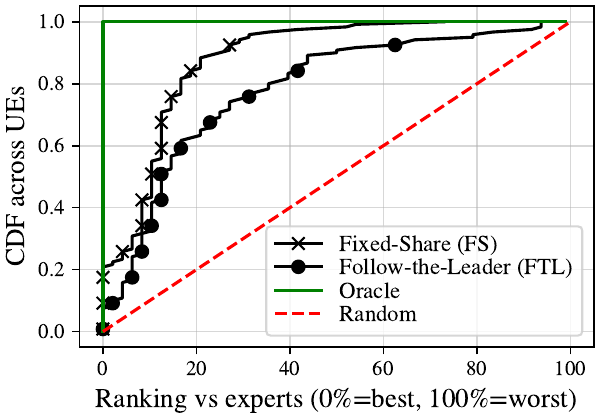}
        \caption{\textbf{Self-tuning capabilities}. Cumulative density function (CDF) of the ranking of expert-advice algorithms vs. single experts in terms of \gls{SINR} estimation accuracy.}
        \label{fig:expert_advice_cdf}
    \end{minipage}\hfill
    \begin{minipage}[t]{0.29\textwidth}
        \centering
        \includegraphics[width=\textwidth]{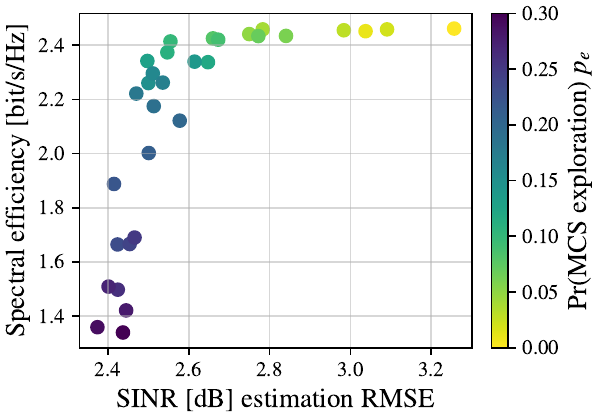}
        \caption{\textbf{Spectral efficiency vs. \Gls{SINR} estimation accuracy} trade-off arising when exploring random \gls{MCS} values with probability $p_e$.}
        \label{fig:se_vs_sinr_est_rmse}
    \end{minipage}
\end{figure*}

\textbf{Comparison with SoA} (Tab.~\ref{tab:sinr_est_cdf_vs_ollas}). We compare the \gls{SINR} estimation accuracy of our method against four \gls{SoA} \gls{LA} schemes: 
i) the industry-standard \gls{OLLA} method \cite{pedersen2007frequency}, which decreases the \gls{SINR} estimate by $\Delta$ upon a NACK and increases it by $\Delta \frac{\tau}{1-\tau}$ after an ACK;
ii) its variant NOLLA \cite{zhu2023nolla}, which adds an exponentially decaying factor to $\Delta$ to reduce OLLA's oscillations; 
iii) \gls{LTS} \cite{saxena2021reinforcement}, which updates its \gls{SINR} estimate via Bayes' rule;
iv) the closest \gls{SoA}, SALAD \cite{wiesmayr2025salad}, whose \gls{SINR} estimation routine is equivalent to our approach for stepsize $\eta=1$, no momentum ($\beta=0$), and no reliance on \gls{CQI} ($\alpha=0$). Our method uses 12 experts, defined on the grid: $\eta\in\{0.5,1,2,3\}$, $\beta\in\{0, 0.15, 0.3\}$. For a fair comparison with \gls{LTS} and SALAD, here we omit \gls{CQI} reports.
For each \gls{UE}, we execute each \gls{SoA} algorithm to produce a \gls{MCS}/ACK/NACK dataset. Then, we apply our method to estimate the \gls{SINR} in an online, open-loop fashion.
Tab.~\ref{tab:sinr_est_cdf_vs_ollas} shows that our method consistently yields more accurate estimates than all considered \gls{SoA} schemes, across all \glspl{UE}.
This suggests that it could serve as a plug-in module for \gls{LA} schemes to benchmark their \gls{SINR} estimation accuracy at run-time and tune their parameters for improved performance.

\textbf{Self-tuning capabilities} (Fig.~\ref{fig:expert_advice_cdf}). We investigate our method's ability to self-tune its parameters $\alpha$ (\gls{CQI} reliance), $\beta$ (momentum), and $\eta$ (stepsize). 
For each \gls{UE}, we consider a dataset containing ACK/NACK feedback, noisy \gls{CQI} reports, generated by adding $\pm 1$ to the true CQI value with equal probability, and \gls{MCS} selected under a \gls{BLER} target $\tau=0.1$. Single expert parameters are picked on the grid: $\alpha\in \{0, 0.1, 1\}$, $\beta\in\{0,0.2,0.4,0.6\}$, and $\eta\in \{0.2,1,1.8,2.6\}$. 
The expert-advice \gls{FS} Alg.~\ref{alg:fixed_share}, with parameters $\mu=5.10^{-4},\epsilon=1$, uses a threshold loss function $\ell_t'(\tgamma_t):=1$ if $|y_t-\bler_t(\tgamma_t)| > 0.5$ and $\ell_t'(\tgamma_t):=0$ otherwise.
In Fig.~\ref{fig:expert_advice_cdf} we compare the \gls{SINR} estimation accuracy of \gls{FS} with that of individual experts. 
\gls{FS} ranks in the top 20\% experts for nearly 90\% of the \glspl{UE}.
In contrast, Follow-the-Leader (FTL) \cite{cesa2006prediction}, which selects the expert with the lowest cumulative loss, underperforms as it fails to track the best expert in non-stationary environments.

\textbf{Spectral efficiency vs. \Gls{SINR} estimation accuracy} (Fig.~\ref{fig:se_vs_sinr_est_rmse}). Although \gls{MCS} selection is outside the scope of this work, we show that using our \gls{SINR} estimate to select the \gls{MCS} induces a trade-off between \gls{SINR} estimation accuracy and spectral efficiency. 
We consider an \gls{LA} scheme that selects in each slot $t$ the highest \gls{MCS} not exceeding an instantaneous \gls{BLER} target $\tau_t$, i.e., $\mcs_t = \max \{ u: \bler(\tgamma_t,\mcs) \le \tau_t \}$. 
The target $\tau_t$ is drawn uniformly in $[0,1]$ with probability $p_e$, and is otherwise computed as $\mathrm{clip} (\tau + 0.02 (\sum_{i<t}\tau - y_t),0,1)$. Note that the feedback term $\sum_{i<t}\tau - y_t$ steers the long-term \gls{BLER} toward $\tau$, while exploratory \gls{MCS} selection probes the channel. 
MCS exploration improves channel estimation accuracy (see Fig.~\ref{fig:se_vs_sinr_est_rmse}) due to the ``surprise'' effect: when the \gls{SINR} abruptly increases (resp., decreases) and a high (resp., low) \gls{MCS} results in an ACK (resp., NACK), the estimate quickly adapts to the new regime. Yet, spectral efficiency drops sharply for $p_e>0.15$. 
This highlights the need for policies that balance exploration and exploitation, probing the channel when beneficial to maximize spectral efficiency, as in \cite{wiesmayr2025salad}.

\section{Conclusions}

This study lays the groundwork for applying the online optimization framework to \gls{SINR}-based resource management. 
Specifically, we formulate \gls{SINR} estimation using ACK/NACK feedback and \gls{CQI} reports from the \gls{UE} within the \gls{OCO} framework (Sec.~\ref{sec:pb_statement}). This new perspective leads to a low-complexity algorithm based on mirror descent with momentum (Sec.~\ref{sec:sinr_est}), for which provable online regret guarantees exist (Sec.~\ref{sec:theory_guarantees}).
Our method is effectively parameter-free, as we leverage the expert-advice framework to automatically tune the algorithm parameters (Sec.~\ref{sec:expert_advice}). This endows the estimator with continual learning capabilities in non-stationary \gls{SINR} environments (Sec.~\ref{sec:theory_guarantees}).
Numerical experiments demonstrate that the proposed approach consistently outperforms \gls{SoA} \gls{LA} schemes in terms of \gls{SINR} estimation accuracy, even with few experts (Sec.~\ref{sec:results}). These results suggest that our method provides a strong benchmark for run-time \gls{SINR} estimation accuracy, and could be used to optimize existing link-adaptation schemes online. 
Moreover, the estimated \gls{SINR} can be directly used to select the \gls{MCS} at each transmission slot. 
In this setting, an exploration-versus-exploitation trade-off naturally arises between refining the \gls{SINR} estimate and exploiting it to maximize spectral efficiency.


\bibliographystyle{IEEEtran}
\bibliography{bib}

\end{document}